\documentclass[fleqn,usenatbib]{mnras}
\usepackage{lmodern}
\usepackage[T1]{fontenc}

\DeclareRobustCommand{\VAN}[3]{#2}
\let\VANthebibliography\thebibliography
\def\thebibliography{\DeclareRobustCommand{\VAN}[3]{##3}\VANthebibliography}

\usepackage{graphicx}
\usepackage{amsmath}	
\usepackage{soul}
\usepackage{ulem}
\usepackage{multirow}
\usepackage{colortbl}
\usepackage[dvipsnames]{xcolor}
\definecolor{tabgreen}{RGB}{44,160,44}
\definecolor{taborange}{RGB}{255,127,14}
\definecolor{tabred}{RGB}{214,39,40}
\definecolor{tabpurple}{RGB}{148,103,189}

\newcommand{\mej}{$M_{\rm{ej}}$}
\newcommand{\vej}{$V_{\rm{ej}}$}
\newcommand{\xlan}{$X_{\rm{lan}}$}
\newcommand{\logmej}{$\log_{10}(M_{\rm{ej}}/M_{\odot})$}
\newcommand{\logvej}{$\log_{10}(V_{\rm{ej}}/c)$}
\newcommand{\logxlan}{$\log_{10}(X_{\rm{lan}})$}
\newcommand{\multinest}{\texttt{PyMultiNest}}

\title[Rapid Kilonovae P.E. Using LFI]{Rapid Parameter Estimation for Kilonovae Using Likelihood-Free Inference}

\author[M. M. Desai et al.]{
M. M. Desai,$^{1,2,3}$\thanks{E-mail: mmdesai@mit.edu (MIT)}
D. Chatterjee,$^{1,2}$
S. Jhawar,$^{5}$
P. Harris,$^{3}$
E. Katsavounidis $^{1,2,3}$
and M. W. Coughlin$^{4}$
\\
$^{1}$ MIT Kavli Institute, 70 Vassar St, MIT, Cambridge, MA 02139, USA \\
$^{2}$ LIGO Laboratory, 185 Albany St, MIT, Cambridge, MA 02139, USA \\
$^{3}$ Department of Physics, MIT, Cambridge, MA 02139, USA \\
$^{4}$ School of Physics and Astronomy, University of Minnesota, 116 Church St SE, Minneapolis, MN 55455, USA \\
$^{5}$ Institut f\"ur Physik und Astronomie, Universit\"at Potsdam, Haus 28, Karl-Liebknecht-Str. 24/25, 14476, Potsdam, Germany
}

\date{Accepted XXX. Received YYY; in original form ZZZ}

\pubyear{\the\year{}}

\begin{document}
\label{firstpage}
\pagerange{\pageref{firstpage}--\pageref{lastpage}}
\maketitle

\begin{abstract}

\noindent 

Rapid parameter estimation is critical when dealing with short lived signals such as kilonovae. We present a parameter estimation algorithm that combines likelihood-free inference with a pre-trained embedding network, optimized to efficiently process kilonova light curves. Our method is capable of retrieving the mass, velocity, and lanthanide fraction of the neutron star ejecta with an accuracy and precision on par with nested sampling methods while taking significantly less computational time. Our inference uniquely utilizes a pre-trained embedding network that marginalizes the time of arrival and the luminosity distance of the signal, allowing inference of signals at distances up to 200 Mpc. We find that including a pre-trained embedding outperforms the use of likelihood-free inference alone, reducing training time, model size, and offering the capability to marginalize over certain nuisance parameters. This framework has been integrated into the publicly available Nuclear Multi-Messenger Astronomy codebase, enabling the broader scientific community to deploy the model for their inference purposes. Our algorithm is broadly applicable to parameterized or simulated light curves of other transient objects, and can be adapted for quick sky localization. 

\end{abstract}


%%%%%%%%%%%%%%%%%%%%%%%%%%%%%%%%%%%%%%%%%%%%%%%%%%

%%%%%%%%%%%%%%%%% BODY OF PAPER %%%%%%%%%%%%%%%%%%

\section{Introduction}

For several decades, the mergers of compact binaries made up of two neutron stars (BNS), or a neutron star with a black hole (NSBH),
have been theorized to emit a sufficient amount of neutron star ejecta to supply heavy elements through the rapid neutron-capture process, commonly referred to as the r-process \citep{1974ApJ...192L.145L, 1989Natur.340..126E, 1999ApJ...525L.121F}. With the observation of the well-studied kilonova AT2017gfo \citep{2017ApJ...848L..27T} and the associated gravitational wave GW170817 and gamma ray burst GRB170817A \citep{PhysRevLett.119.161101}, recent research has been done on estimating the properties of the neutron star ejecta \citep{2017Natur.551...80K, 2023MNRAS.520.2558B}. This has led to developments in understanding the properties and equation of state of neutron stars \citep{2018ApJ...852L..29R, 2024MNRAS.52711053M}, additional measurements of the Hubble constant \citep{2022Univ....8..289B}, and investigating the contributions of these mergers to heavy element production via the r-process \citep{2019EPJA...55..203S}. 

Current operational detectors such as the Zwicky Transient Facility (ZTF, \cite{ztf}) and the Wide-Field Infrared Transient Explorer (WINTER, \cite{2022ApJ...926..152F}) are optimized to target electromagnetic (EM) counterparts to gravitational waves. Additionally, the upcoming Vera C. Rubin Observatory (previously referred to as Large Synoptic Survey Telescope, LSST) will have the opportunity to detect kilonovae signals \citep{2019ApJ...873..111I, 2019ApJ...874...88C, 2019PASP..131f8004A, 2022ApJS..258....5A, 2024APh...15502904A}. Improvements to current gravitational-wave detectors and plans to construct next-generation detectors are expected to lead to an increase in known BNS/NSBH events and lead to an order of 10 new kilonova observations per year with the Vera C. Rubin Observatory and the Nancy Grace Roman Space Telescope \citep{2023PhRvD.107l4007G}. 

In anticipation of these events, an effort is being made to create models to simulate potential light curves. The first simple kilonova model was developed by \cite{1998ApJ...507L..59L}, which assumed an adiabatically expanding spherically symmetric envelope. Solutions to their equations provided an estimate of the luminosity and duration of a kilonova. Subsequently, the observation of GW170917 led to a two-part framework: an initial lanthanide-poor blue emission, and a secondary lanthanide-rich red emission \citep{2017LRR....20....3M}. Several models have been developed to understand various aspects of kilonova \citep{2017Natur.551...80K, 2020Sci...370.1450D, 2021NatAs...5...46A, 2021ApJ...906...94Z, 2023arXiv230711080A, 2021ApJ...918...10W, 2023PhRvR...5a3168K, 2023MNRAS.526.4585G}, some of which are derived utilizing the three-dimensional Monte Carlo radiation transport modelling code POSSIS \citep{2019MNRAS.489.5037B}. To consolidate these models, a nuclear multi-messenger astronomy package referred to as NMMA offers tools to generate light curves from several models and perform parameter estimation on these simulated events \citep{Pang:2022rzc}. We choose the spectral energy distribution (SED) model grid from \cite{2017Natur.551...80K}, hereby referred to as Ka2017, for our light curve generation through NMMA. The implementation of this model is further discussed in Section~\ref{Data}.

Over the past few years, developments in machine learning (ML) techniques have explored various avenues in reducing the computational cost and time for studying light curves. While ML-based improvements have been made in the vitesse of light curve generation using variational autoencoders \citep{2024ApJ...961..165S} and neural networks \citep{2024arXiv240205871P}, likelihood-free parameter estimation of this data remains an untapped field. A prohibitive step of sampling methods is the likelihood calculation, $p(d|\theta)$. To avoid this calculation, likelihood-free or simulation-based inference (LFI/SBI) offers a potential alternative method (see \cite{2020PNAS..11730055C} for a review). 

In this paper, we describe our framework for faster parameter inference using LFI with comparable accuracy and precision to nested sampling \citep{10.1214/06-BA127}. We simulate light curves using the Ka2017 model grid and ZTF noise realizations. An embedding neural network marginalizes the distance to the source and the exact peak time over a data segment. This allows us to focus on the intrinsic parameters of the light curve such as the ejecta mass ({\mej}), velocity ({\vej}), and lanthanide fraction ({\xlan}), that govern the light curve's appearance. Further discussion of our choice of priors and simulation procedure are described in Section~\ref{Data}. We achieve marginalization by creating alternate, or shifted, views of each light curve with the same intrinsic parameters, while varying the time and luminosity distance. These views are jointly embedded by minimizing the variance, invariance, and covariance of the two data views. Conditioned on this data summary, a normalizing flow is used to infer these kilonova parameters on a logarithmic scale: \logmej, \logvej, and \logxlan. We compare our posterior distributions to nested sampling results produced with the {\multinest} package \citep{2009MNRAS.398.1601F}. Section~\ref{Method} details the ML architecture and demonstrates its efficacy on simulated kilonova light curves, with the results provided in Section~\ref{Results}.

\section{Data} \label{Data}

We simulate light curve data using the Ka2017 radiative transfer method, which is available through the NMMA package \citep{2017Natur.551...80K, 2024ascl.soft02001P}. We focus on obtaining realistic light curves viewable from ZTF, thereby simulating data in all three of its bands: \texttt{ztfg}, \texttt{ztfr}, and \texttt{ztfi}. 

The composition and appearance of a kilonova is controlled by the spectral energy distribution (SED). The Ka2017 model parametrizes this SED through the combined dynamical and disk mass (\mej), velocity (\vej), and lanthanide fraction (\xlan) of the ejecta. Lanthanide fraction, \xlan, can serve as a metric for heavy r-process matter, as lanthanides contribute anywhere from 1 to 10 \% of the total ejecta mass \citep{2017Natur.551...80K}. The line transitions from lanthanides greatly increases the opacity, resulting in long diffusion times and longer light curves. The Ka2017 model accounts for two kilonova mass ejection methods: a brief stripping of the stars due to tidal forces ($10^{-3} M_\odot < M_{\rm{ej}} < 10^{-2} M_\odot$), and a secondary ejection of mass from the accretion disk blown by neutrino-driven winds ($10^{-2} M_\odot < M_{\rm{ej}} < 10^{-1} M_\odot$) \citep{2015ApJ...813....2M}. The initial shock occurs at high speeds on the order of a dynamical velocity $V_{\rm{dyn}} \sim 0.2c - 0.3c$, while the disk mass disperses at rates of $V_{\rm{disk}} \sim 0.05c - 0.1c$ \citep{2017Natur.551...80K}. These quantities also affect the peak of the light curve: larger ejecta masses produce brighter and longer-lasting light curves, and faster ejecta velocities produce brighter yet short-lived light curves. Ka2017 combines these quantities as the ejecta parameters \mej\ and \vej, which we use to parametrize our model. In this study, we neglect the effects of host galaxy and Milky Way extinction effects on the light curve.

\begin{table}
	\centering
	\caption{Prior boundaries for data generation.}
	\label{tab:dataprior}
	\begin{tabular}{lccr} % four columns, alignment for each
		\hline\hline
		Parameter & Minimum & Maximum \\
		\hline
            \logmej & -1.9 & -1 \\
            \logvej & -1.52 & -0.53 \\
            \logxlan & -9 & -3 \\
            $t$ (days) & -2 & 6 \\
            $d_L$ (Mpc) & 50 & 200 \\
		\hline\hline
	\end{tabular}
\end{table}

The investigation of kilonovae can be divided into two problems: search and inference. Here, we assume that ZTF is capable of discovering optical counterparts of gravitational waves to distances of up to 200 Mpc, and we simulate our light curves accordingly \citep{2021ApJ...918...63A}.
To ensure that light curves are visible by ZTF, we selected broad parameter ranges that were capable of producing detections in addition to the considerations provided by \cite{2017Natur.551...80K}. These priors are detailed in Table~\ref{tab:dataprior}. Prior to training the model, all five parameters are normalized between 0 and 1. We choose time and luminosity distance, $(t, d_L)$, as our nuisance parameters that we wish to marginalize. Our time prior is selected to create a conservative variability for where the light curve peaks. Our distance prior allows our model to be sensitive to the edge of EM detectablility with ZTF and the expected two detector LIGO O4 BNS inspiral range of 25-130 Mpc \citep{2018LRR....21....3A}. While GW170817 was detected at 40 Mpc, optical surveys are expected to be sensitive to kilonovae at a distance of up to 200 Mpc with a limiting magnitude of 22 mag \citep{2021MNRAS.504.1294S}.

Our data is simulated at a regular cadence of one detection every six hours over a 20 day period of time. While this is an optimistic detection scheme, we discuss potential relaxation schemes in Section~\ref{Results}. We sample uniformly from our intrinsic parameters \logmej, \logvej, and \logxlan, and our extrinsic parameters $t$ and $d_L$. In the case where the light curve is dim or short-lived, non-detections are encoded as the magnitude detection limit of 22.0 in all bands. This is a typical conservative restriction given that ZTF is limited to 20.8, 20.6, and 19.9 mag [21.6, 21.4, and 21.9 mag] for the g, r, and i bands for a 30s [300s] exposure time \citep{2019PASP..131a8002B, 2021MNRAS.504.2822A}. Since the broad priors produce both extremely short and long light curves, we pad the data with our non-detection limit in order to maintain a standard length of data for our model. For training purposes, a minimum amount of 8 cumulative detections across the three bands is set as a requirement. 

Additionally, each light curve is simulated with unique instrument noise to ensure that all views of the same intrinsic parameters are distinct. These noise realizations come from the three noise distributions of the ztfg, ztfr, and ztfi bands. The noise instances are sorted into 12., 18., 20., 21., and 23 magnitude bins to obtain an accurate noise sampling for each data point in the light curve. The final processing step is a normalization performed across our training and test data. Neglecting the non-detections, we compute the mean and standard deviation of our training sample and apply the following procedure to normalize the photometric data:

\begin{equation}
    z_i = \frac{x_i - \bar{x}}{\sigma},
\end{equation}

\noindent where $x_i$ represents our original data and $z_i$ the new normalized photometry.

To train the network, we create a set of ``fixed'' data, $d$, which share the same initial arrival time $t=0$ and a luminosity distance of $d_L = 50$ Mpc. For each fixed light curve, an alternative ``shifted'' light curve $d'$ is generated by modifying $t$ and $d_L$ according to the priors in Table~\ref{tab:dataprior}. A single fixed light curve will have its shifted counterpart that shares the precise values for \logmej, \logvej, and \logxlan, resulting in the same physical phenomena but a different observed light curve. In this way, we create two views of the data generated by the same intrinsic parameters with distinct noise instances. We utilize both views to train our data summary, and only the shifted view to train our normalizing flow.

\section{Method} \label{Method}

In this section, we present our ML-based framework for parameter estimation using LFI. We introduce the concept of an embedding network that permits the marginalization of certain parameters and reduces the size of the following inference network in Section~\ref{Data Summary}. The embedding also serves to reduce the high-dimensional light curves, each consisting of 121 points multiplied by three bands, to a 7 dimensional representation. The pre-trained embedding is supplied to a normalizing flow, which performs the parameter inference as described in Section~\ref{Norm Flow}. 

\subsection{Data Summary} \label{Data Summary}

\begin{figure*}
    \centering
    \includegraphics[width = 15 cm]{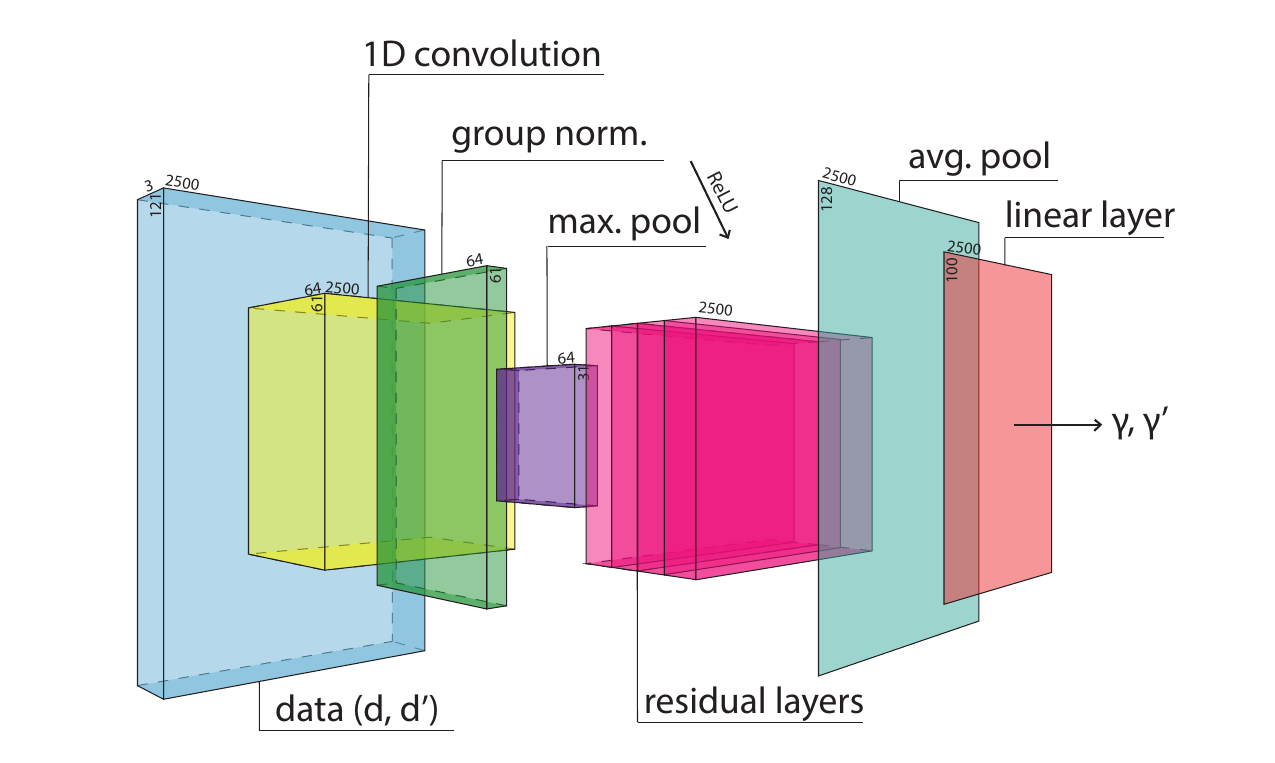}
    \caption{Diagram of the ResNet architecture used for our contraction network, $f$. We utilize the original light curve data, d, and an augmented version, d', where the time of arrival and luminosity distance are shifted according to the priors in Table~\ref{tab:dataprior}. Data d and d' are separately processed through $f$ to obtain $\gamma$ and $\gamma'$ as the initial step of our embedding network $\Gamma$. This enables a mapping of physical parameters while marginalizing time and distance.}
    \label{fig:resnetarch}
\end{figure*}

In order to perform LFI using normalizing flows, we sample parameters $\boldsymbol{\theta}_i \sim p(\boldsymbol{\theta})$, and simulate light curves $\boldsymbol{d}_i \sim p(\boldsymbol{d} \mid \boldsymbol{\theta}_i)$. The resulting pairs $\{\boldsymbol{\theta}_i, \boldsymbol{d}_i\}$ are used to learn an approximation of the posterior distribution, $q_\phi(\boldsymbol{\theta} \mid \boldsymbol{d})$, by maximizing the log-likelihood,

\begin{equation}
    \log \mathcal{L}(\phi) = \sum_i \log q_\phi(\boldsymbol{\theta}_i \mid \boldsymbol{d}_i).
    \label{loglikelihood}
\end{equation}

\noindent where $q_{(\phi)}$ is the model's estimated posterior.

A key challenge in implementing LFI on our dataset is the high dimensionality of our light curves. To combat this, we train an effective data summary that serves a dual purpose: reducing data dimensionality and marginalizing nuisance parameters: time and distance. In this case, we are primarily interested in the intrinsic parameters, like \mej, \vej, or \xlan, though the data depends on other extrinsic parameters. Kilonova signals can be seen as unique if they arrive at different times or occur at different distances, even when the intrinsic parameters of the ejecta are the same. The recorded peak time of a kilonova depends not only on its arrival time and luminosity distance, but also on the underlying model, providing yet another reason to marginalize over it. An LFI network unaware of these translations requires the extrinsic parameters of time and distance to be inferred alongside the other physical parameters or marginalized over, leading to a larger, more expensive network. To mitigate the size of the network, we can marginalize the nuisance parameters by creating an appropriate data summary that preserves the information of only the intrinsic parameters and learns the symmetric patterns of $t$ and $d_L$ through different views of the data. We utilize self-supervised learning (SSL) to accomplish this goal.

This technique has been used previously in \cite{2023arXiv231207615C} for simpler, lower-dimensional signal morphologies, demonstrating that the creation of an embedding space aids in faster parameter inference. The application of self-supervised learning to create a summary extends to compressing cosmological data and encoding symmetries of partial differential equations as seen in \cite{2024MNRAS.527.7459A, 2023arXiv230207842M}. Additionally, self-supervised learning has been used to simplify the complexity of gravitational waves as an initial denoising step \citep{2024arXiv240304350L}. \cite{2020arXiv201202807R} demonstrates a different approach (YuleNet) for data summarization for LFI with timeseries data. As a result, we also implement a summarization step for our high-dimensional light curve data. 

\begin{figure}
    \centering
    \includegraphics[width = \columnwidth]{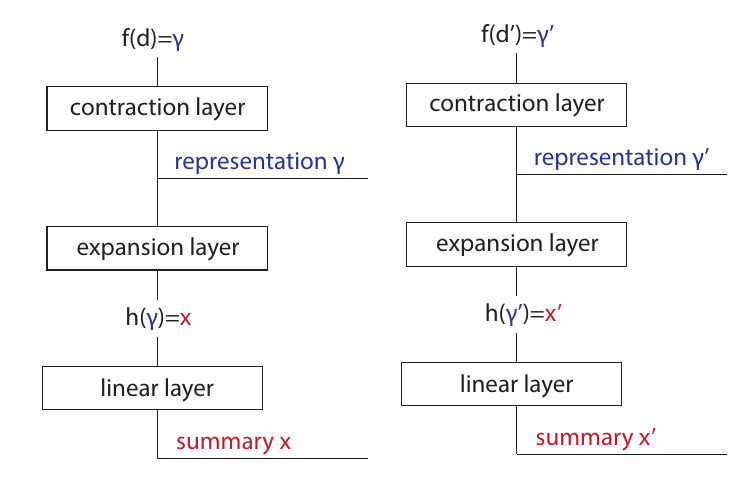}
    \caption{Depiction of our full data summary network. Our data, $d$ and $d'$ are initially passed through the ResNet $f$. After a contraction layer, we extract the representations $\gamma$ and $\gamma'$. The representation is expanded and passed through the $h$ layer, which is comprised of a linear layer. The output is passed through a final linear layer to output the data summaries, $x$ and $x'$.}
    \label{fig:embeddingarch}
\end{figure}

We create a reduced representation where data with similar intrinsic parameters but different extrinsic parameters are represented in a joint embedding space. This is done by projecting both fixed and shifted data into a 7-dimensional embedding space via a neural network, 

\begin{equation}
    \label{eqn:net}
    \Gamma \equiv h \circ f,
\end{equation}

\noindent where $f$ is an adapted ResNet \citep{2015arXiv151203385H}, and $h$ is a fully-connected expander layer. The architecture of $f$ is mapped in Figure~\ref{fig:resnetarch}. Data from the three photometric bands are passed as 3 separate channels to $f$. We utilize the modifications in \cite{2024arXiv240318661M} to alter the original ResNet54 architecture, namely the use of 1D convolutions and group normalization in opposition to 2D convolutions and batch normalization. Group normalization acts along the channel dimension, allowing data from all three photometric bands to be processed together. Because the heavy element composition of these light curves greatly influences the color relationship between photometric bands, we prefer the use of group normalization to preserve the color information. We denote this projected space as the data \emph{representation}. The dimensionality of the resulting representation is determined to be 7 through hyperparameter tuning, which is detailed in Section~\ref{Tuning} and Appendix~\ref{app:tuning}.

Our full network architecture is explained in Figure~\ref{fig:embeddingarch}'s flowchart. After the representation has been extracted, we pass $\gamma$ and $\gamma'$ to our 150-dimensional expander layer, $h$, such that the outputs $x, x'$ are produced:

\begin{eqnarray}
    \label{eqn:rep}
    \gamma = f(d) &,& \gamma' = f(d'), \nonumber \\
    x = h(\gamma) &,& x' = h(\gamma').
\end{eqnarray}

\noindent Here, $x$ and $x'$ represent the data in an embedding space, which we denote as the data \emph{summary}. Our performance hinges on a sizable $h(\gamma)$. Implementing a large, fully-connected linear layer allows information distributed across several nodes to flow between one another. This enables the network to condense information across the entire light curve.

To correlate $x$ and $x'$, we utilize the VICReg loss function~\citep{2021arXiv210504906B}, which is calculated as a weighted sum on batches $X = [x_{1}, ..., x_{n}]$ and $X' = [x_{1}', ..., x_{n}'] $:

\begin{equation}
\label{eqn:loss}
\begin{split}
\mathcal{L}_{VICReg}(X, X') & = \lambda_1 MSE(X, X') + \lambda_2[\sqrt{Var(X) + \epsilon} \\
& + \sqrt{Var(X') + \epsilon}] + \lambda_3[C(X) + C(X')],
\end{split}
\end{equation}

\noindent where $C(X)$ is the covariance matrix, defined by:

\begin{equation}
    C(X) = \frac{1}{n-1}\sum^{n}_{i=1}(x_i - \bar{x})(x_i - \bar{x})^{T}
\end{equation}

\noindent where

\begin{equation}
    \bar{x} = \frac{1}{n}\sum^{n}_{i=1}x_{i}.
\end{equation}

We choose equal weights $\lambda_1 = \lambda_2 = \lambda_3 = 1$ for this study. We minimize the VICReg similarity loss between the embeddings $x$ and $x'$ such that the regularized variance and covariance of the batches are individually reduced, while the invariance between batches is reduced through mean-squared error. 10 minutes of processing time is required to train the network from scratch for 50 epochs on one NVidia V100 GPU.

To properly encode the effects of changing $t$ and $d_L$, we must create two views of the data. We repetitively create 50 light curves using one identical combination of our physical parameters (\logmej, \logvej, and \logxlan) with unique noise instances. This is done iteratively with novel combinations of the physical parameters to produce a dataset. Each of light curves in this fixed set are positioned at a $d_L$ of 50 Mpc, and start at the same time. Our second data view is created by adjusting $t$ and $d_L$ of each light curve of the fixed set according to the priors listed in Table~\ref{tab:dataprior}. Our training dataset contains 8,729 unique combinations of {\mej}, {\vej}, and {\xlan}, resulting in 872,900 total light curves. The data representation, $\gamma$, is trained jointly using these sets of fixed and shifted light curves.

\begin{figure*}
\centering
\includegraphics[width = 8.5 cm]{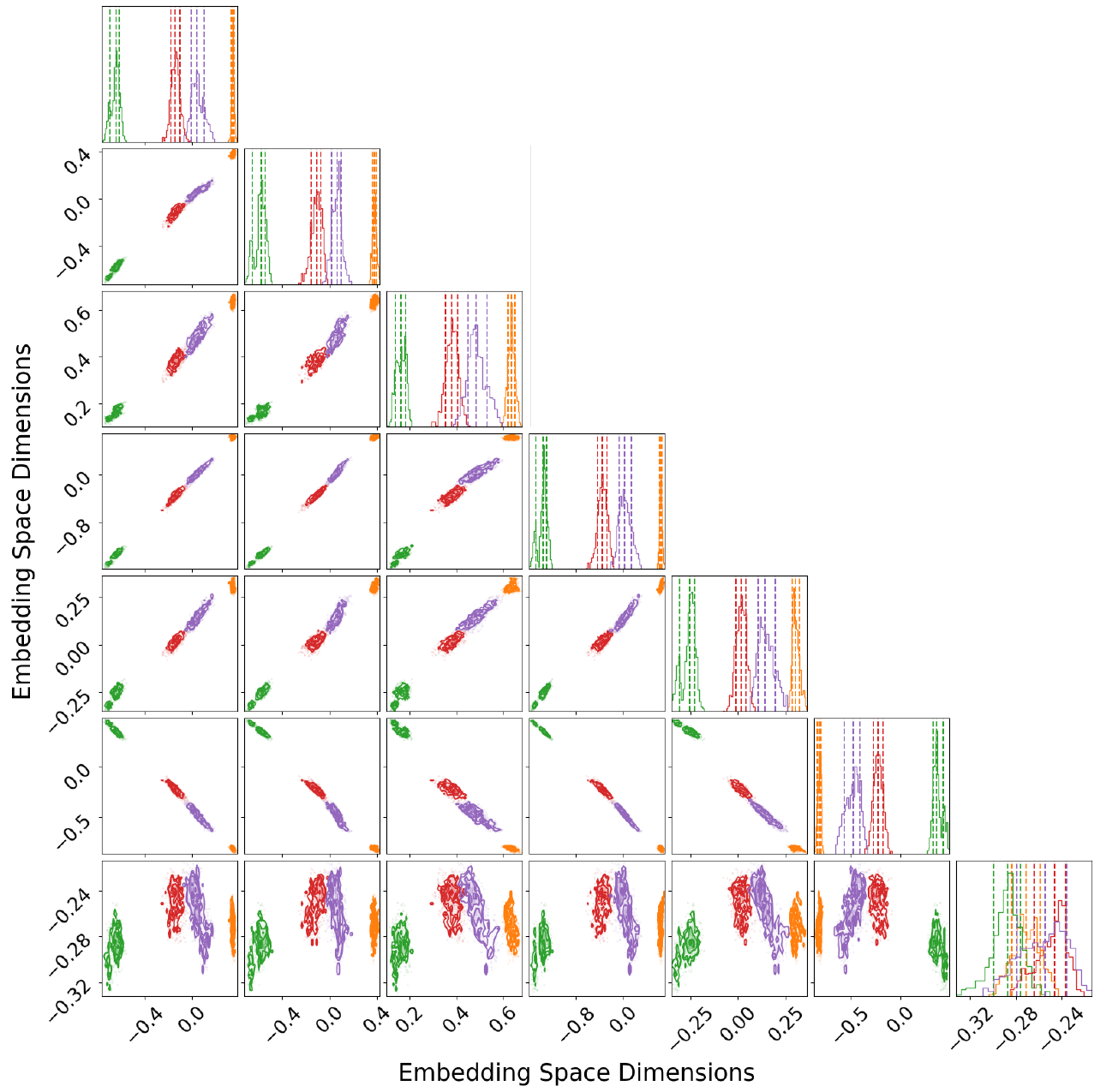}
\includegraphics[width = 8.5 cm]{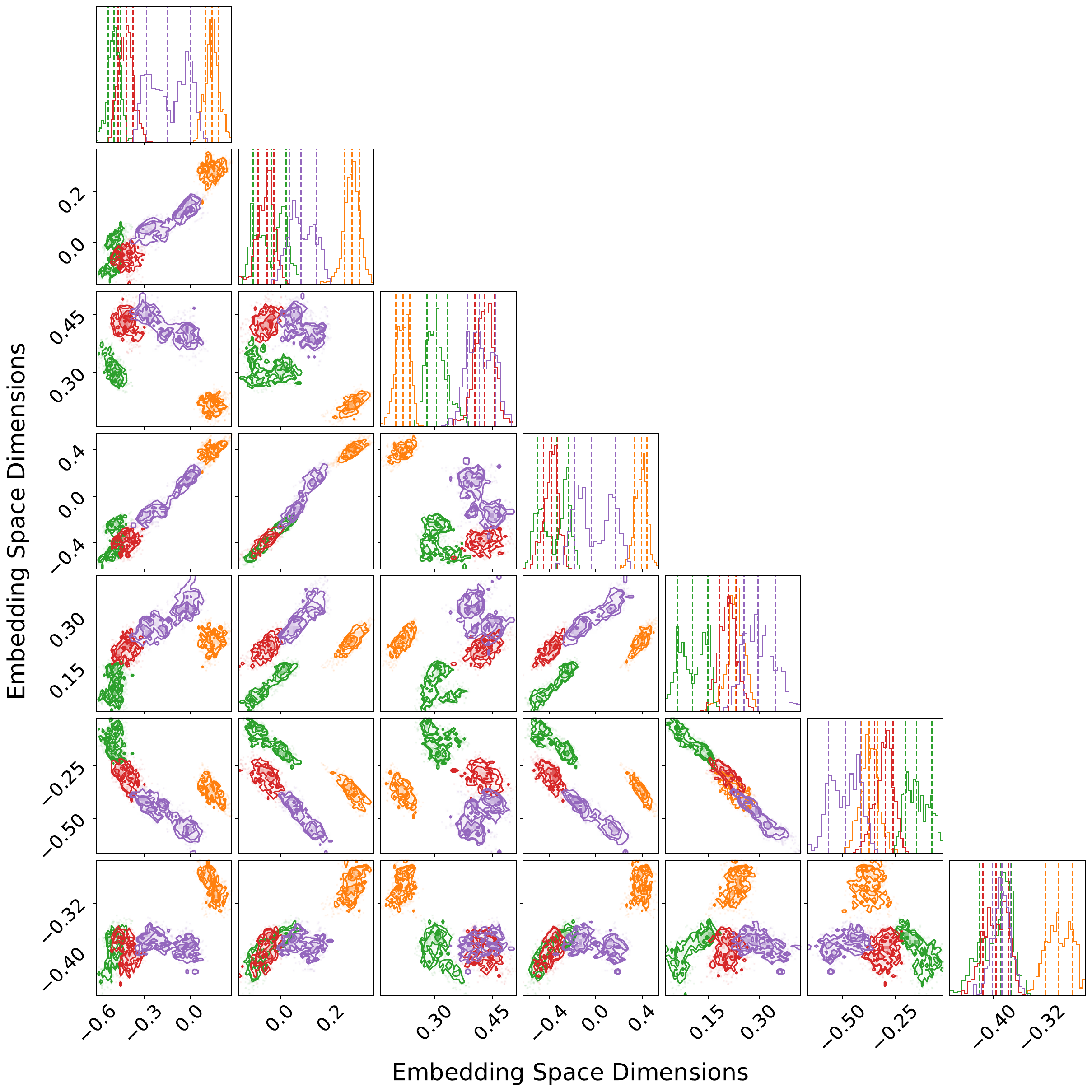}
\caption{In this figure, we see four combinations of \logmej, \logvej, and \logxlan\ plotted in orange, green, red, and purple. Each set contains 1000 light curves that have been shifted in time and distance. These plots represent a data summary space. The left plot displays a pre-trained embedding, and the right plot displays the embedding after it has continued training alongside the normalizing flow. By passing an informed embedding where parameters have been clustered and allowing it to further be trained, we find that the embedding space develops a more complex mapping to further separate different light curves while maintaining clustering of distinct \logmej, \logvej, and \logxlan.}
\label{fig:embedding}
\end{figure*}

\begin{table*}
\centering
\renewcommand*{\arraystretch}{1.5}
\begin{tabular}{lccc}
\hline\hline
& $\log_{10}(M_{\text{ej}})$ & $\log_{10}(V_{\text{ej}})$ & $\log_{10}(X_{\text{lan}})$ \\ \hline
\cellcolor{taborange} Set 1 & $-1.00 > M > -1.25$ & $-0.53 > V > -0.75$ & $-3.00 > X > -4.50$ \\
\cellcolor{tabgreen} Set 2 & $-1.25 > M > -1.50$ & $-0.75 > V > -1.00$ & $-4.50 > X > -6.00$ \\
\cellcolor{tabred} Set 3 & $-1.50 > M > -1.75$ & $-1.00 > V > -1.25$ & $-6.00 > X > -7.50$ \\
\cellcolor{tabpurple} Set 4 & $-1.75 > M > -1.90$ & $-1.25 > V > -1.52$ & $-7.50 > X > -9.00$ \\ \hline\hline
\end{tabular}
\caption{Parameter ranges for the four ejecta clusters shown in the plot above. Each set corresponds to a distinct injection of \logmej, \logvej, and \logxlan, with rows color-coded to match their respective cluster colors in the plot.}
\end{table*}

Figure \ref{fig:embedding} displays the representation $\gamma'$ for four combinations of physical injection parameters to showcase various areas of the prior selection. A visual comparison of the left and right plots highlights the importance of allowing a portion of the embedding weights to remain trainable alongside the normalizing flow. We observe that after training, light curves of different \logmej, \logvej, and \logxlan are clustered even with different values of $t$ and $d_L$, indicating that the two invariant parameters have been successfully marginalized. 

\subsection{Normalizing Flows} \label{Norm Flow}

For our LFI, we leverage normalizing flows, which transform a simple initial density into a complex distribution by composing a series of learnable transformations \citep{2015arXiv150505770J, nflows}. We use a three-dimensional standard normal base distribution and Masked Affine Autoregressive Transforms to construct the flow, as this transformation allows efficient, single-pass density evaluation \citep{2017arXiv170507057P}. The flow is conditioned on the \emph{representation}, $f(d)$, from Eq.~(\ref{eqn:rep}) which summarizes the input light curves as described in Section~\ref{Data Summary}. The model is trained by minimizing the \textit{negative} log-likelihood (i.e., $-\log \mathcal{L}(\phi)$ from Eq.(~\ref{loglikelihood})) using the Adam optimizer. Additionally, we halve the learning rate when the loss plateaus with a patience of 2 epochs to avoid stagnation. 

Only the shifted view is utilized to train the normalizing flow. 472,300 shifted light curves with unique noise and unique combinations of the 5 parameters of Table~\ref{tab:dataprior} comprise the training dataset. Hence, this method is suitable for the realistic scenario where there is uncertainty in the arrival time or the distance to the source is not known precisely. The representation $f(d)$, ensures that this uncertainty is marginalized over. Our best performing model is composed of 9 blocks, 5 transforms, and 90 hidden layers, informed by a hyperparameter search (see Section \ref{Tuning}). 

\begin{figure} 
\centering
\includegraphics[width = \columnwidth]{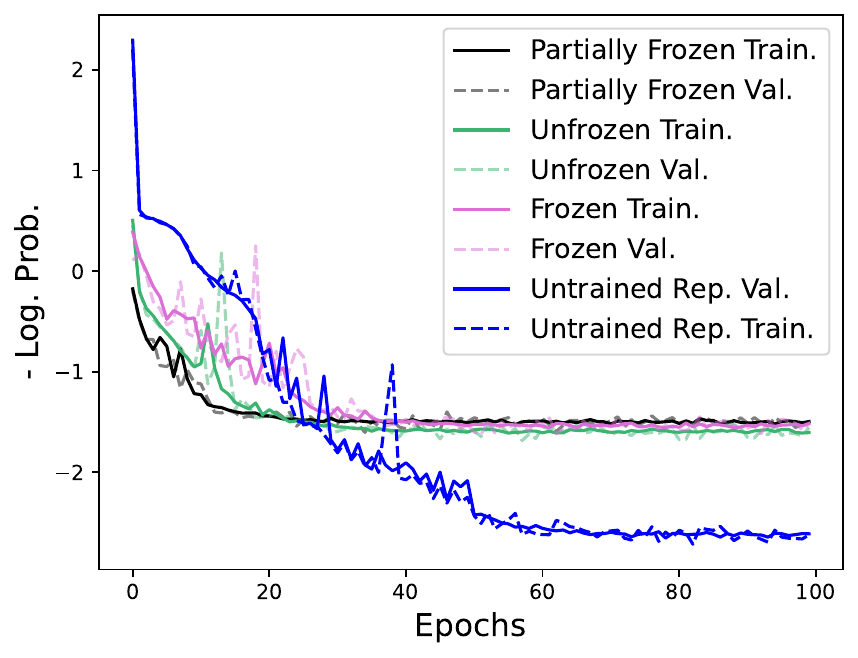}
\includegraphics[width = \columnwidth]{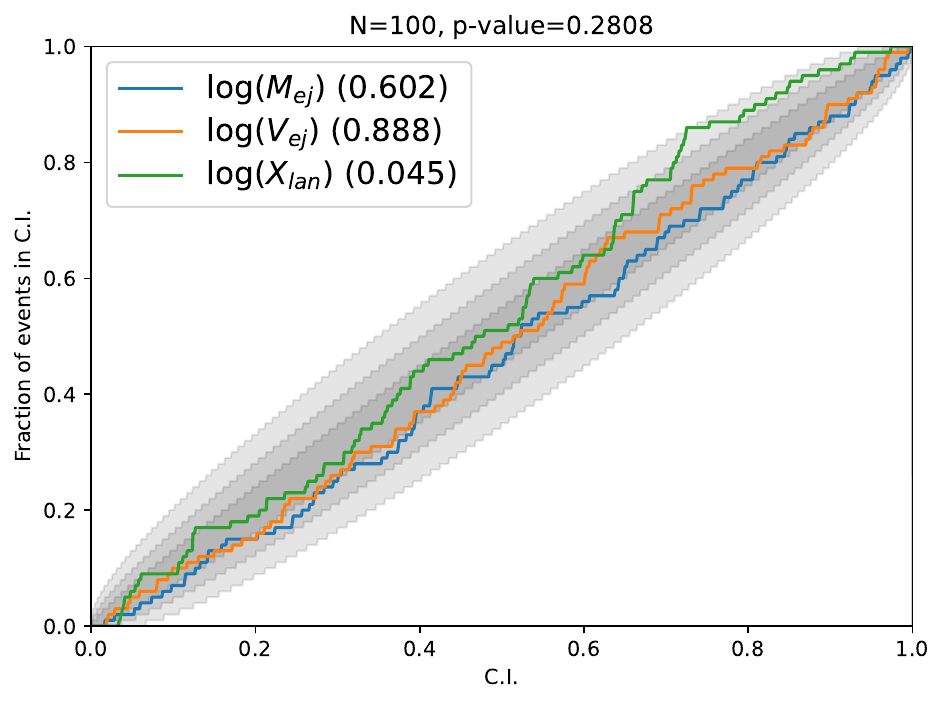}
\caption{\textbf{Upper:} Training (solid) and validation (dashed) loss curves  for variations of the flexibility of the embedding network. The use of an untrained representation is shown in blue. The pink, black, and green curves are pretrained embeddings with frozen, partially frozen ($f(d)$ flexible), and unfrozen parameters ($f(d)$ and $h(d)$ flexible), respectively. We find that allowing some flexibility of the representation network results in the fastest convergence and the most accurate parameter estimation. \textbf{Lower:} Percentile-percentile plots from 100 test light curves, each with 1000 posterior samples drawn from the flow. The fraction of instances where the parameters lie within the confidence interval is plotted for the three inferred parameters, with \logmej in blue, \logvej in orange, and \logxlan in green. Overlaid are the $1 \sigma$, $2 \sigma$, and $3 \sigma$ confidence intervals in decreasing opacity. We find that our samples lie within the confidence intervals, indicating that the correct true parameters are inferred consistently.}
\label{fig:flowloss}
\end{figure}

The loss curves displayed in the upper panel of Fig.~\ref{fig:flowloss} demonstrate the negative log probability of several embedding states, all utilizing a fixed batch size of 25, learning rate of $9.2\times10^{-5}$, and an identical training dataset. The learning rate, batch size, and configurable model parameters are determined through hyperparameter tuning. See Section~\ref{Tuning} and Appendix~\ref{app:tuning}. To demonstrate the efficacy of the data summary, we track the effects of several variations of the summary informing our flow. In the frozen case, we precondition the embedding but freeze all trainable weights when it is relayed to the normalizing flow. The unfrozen case allows all parameters of the embedding to vary with the flow. For the partially frozen case, we only freeze the weights of the $h$ layer, leaving $f(d)$ flexible. From Fig.~\ref{fig:embedding}, we confirm that allowing partial flexibility enables further clustering of parameters and directly correlates to an increase in our estimation accuracy. 

The upper panel of Fig.~\ref{fig:flowloss} further showcases the difference in convergence times of the various embedding states. We find that the partially frozen configuration converges the fastest, allowing us to select a lighter model with approximately 1 million configurable parameters. On the other hand, an untrained representation initialized with random weights converges slower. Though the untrained representation is capable of achieving lower losses, we find that the model returns uninformative posteriors and cannot correlate two light curves with similar intrinsic parameters. The trained embedding purposely maintains the structure of the latent space, preventing a further decrease in loss. We conclude that utlizing a data summary step is valuable in marginalizing the time and distance values and reducing the overall size of the model. For a model size of $\sim$~1.2 million parameters, training requires 50 minutes for 100 epochs on one NVidia V100 GPU.

We reserve a small sample of 100 light curves for testing the flow and {\multinest}. 1000 posteriors are drawn from the trained flow for each light curve, and we calculate fraction of events that lie within the credible interval. This is plotted in the lower panel of Figure~\ref{fig:flowloss}, where we see that the test cases all lie within the credible $3 \sigma$ bands. Our model consistently returns accurate inferences for all three parameters. Additionally, diagonal lines indicate that the model has high confidence in the results it returns. Section~\ref{Results} further discusses the resulting posteriors from {\multinest} and the flow.

\subsection{Hyperparameter Tuning} \label{Tuning}

To achieve optimal configurations of network parameters, hyperparamenter tuning is implemented. We utilize the \texttt{ray.tune} Python package to determine batch sizes and learning rates for our training, as well as find ideal configurations of our model parameters \citep{liaw2018tune}. We searched over 100 combinations of potential dimensionality, kernel sizes, number of $h$ layers, number of blocks, and number of final dimensions for our representation network $\Gamma$ for a maximum of 50 epochs, or steps. Similarly, we searched through 100 combinations of number of transforms, blocks, and hidden features for the normalizing flow over a maximum of 100 steps. The choices we used are displayed in Table~\ref{tab:config}, where options in square brackets are discrete and those in parenthesis are continuous. We use the Async HyperBand Scheduler from \texttt{Ray} to terminate unpromising trials in an effort to reduce the required GPU time, and the best run is selected based on the lowest average validation loss achieved. We utilize the informed tuning to finalize our selection of model parameters. See Appendix~\ref{app:tuning} for configuration specifics and loss metrics for the top ten hyperparameter trials. We also note that lower-dimensional models generally perform with a decreased accuracy and precision. In particular, we find that reducing the dimensionality of the embedding network disrupts the adjustments made by the normalizing flow, resulting in a preferred higher dimension of 7.

\begin{table}
        \centering
	\caption{We select parameter spaces over which we use \texttt{Ray} to search for the best configuration for our models. Below shows the possible options for the representation network and for the flow.}
	\label{tab:config}
	\begin{tabular}{lll} 
		\hline\hline
		Model & Parameter & Options \\
		\hline
             & Dimensions & [3, 4, 5, 7, 8] \\
             & Kernel Size & [5, 7, 9, 13] \\
             & $h$ Layers & [1, 2, 3] \\
            Representation & Blocks & [3, 4, 5, 6] \\
             & Final Dimension & [2, 5, 10] \\
             & Batch Size & [25, 50, 75] \\
             & Learning Rate & LogUniform($10^{-6}$, $10^{-4}$) \\
             \hline
             & Blocks & [4, 5, 6, 7, 8, 9] \\
             & Transforms & [5, 6, 7, 8, 9, 10] \\
             Flow & Hidden Features & [60, 70, 80, 90, 100] \\
             & Batch Size & [25, 50, 100] \\
             & Learning Rate & LogUniform($10^{-6}$, $10^{-4}$) \\
		\hline\hline
	\end{tabular}
\end{table}

\section{Results} \label{Results}

Our parameter estimation posteriors are shown in comparison to {\multinest} inferences in Section~\ref{posteriors}. We also investigate the effect of missing data in Section~\ref{missingdata}, with a continued investigation into a slower cadence scheme in Appendix~\ref{app:resilience}. The inference for each pair of fixed and shifted light curves are in agreement with each other and competitive with results through nested sampling. 

\subsection{Posteriors} \label{posteriors}

\begin{figure*}
\centering
\includegraphics[width = 8 cm]{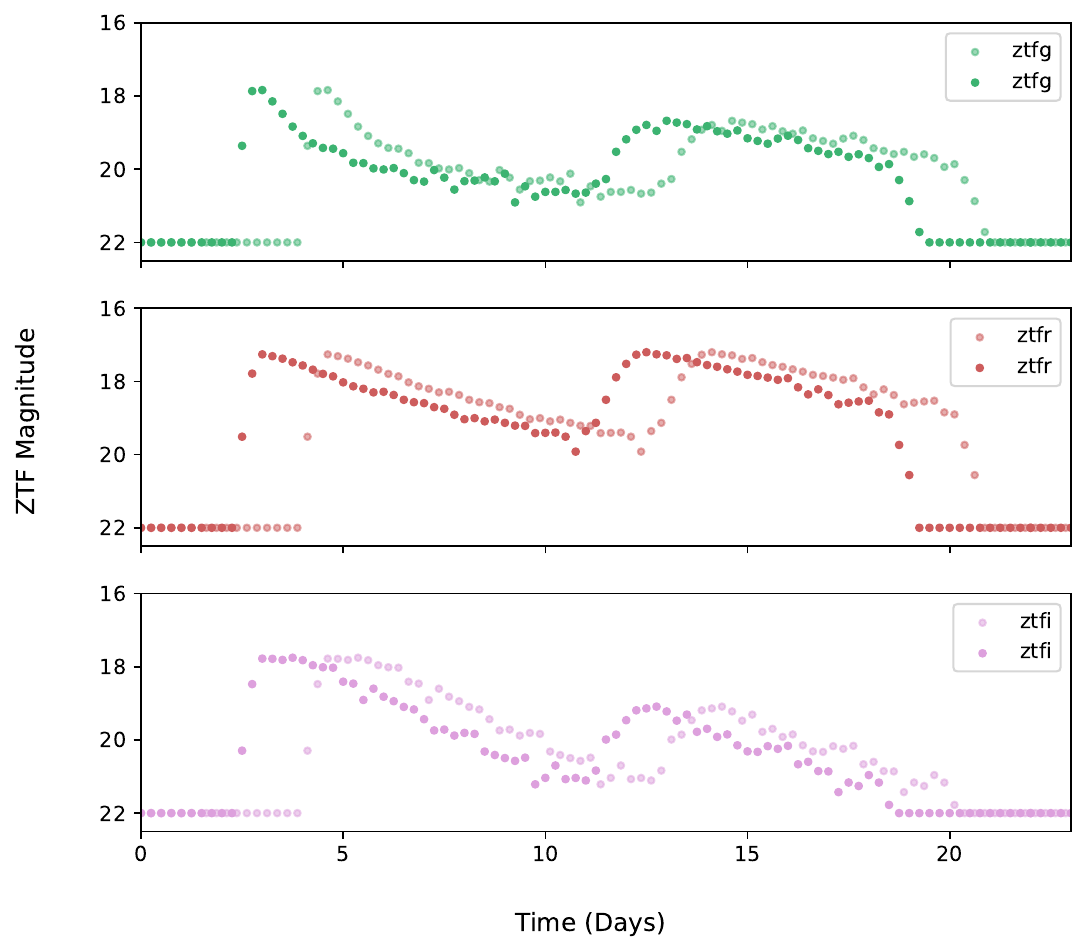}
\includegraphics[width = 8 cm]{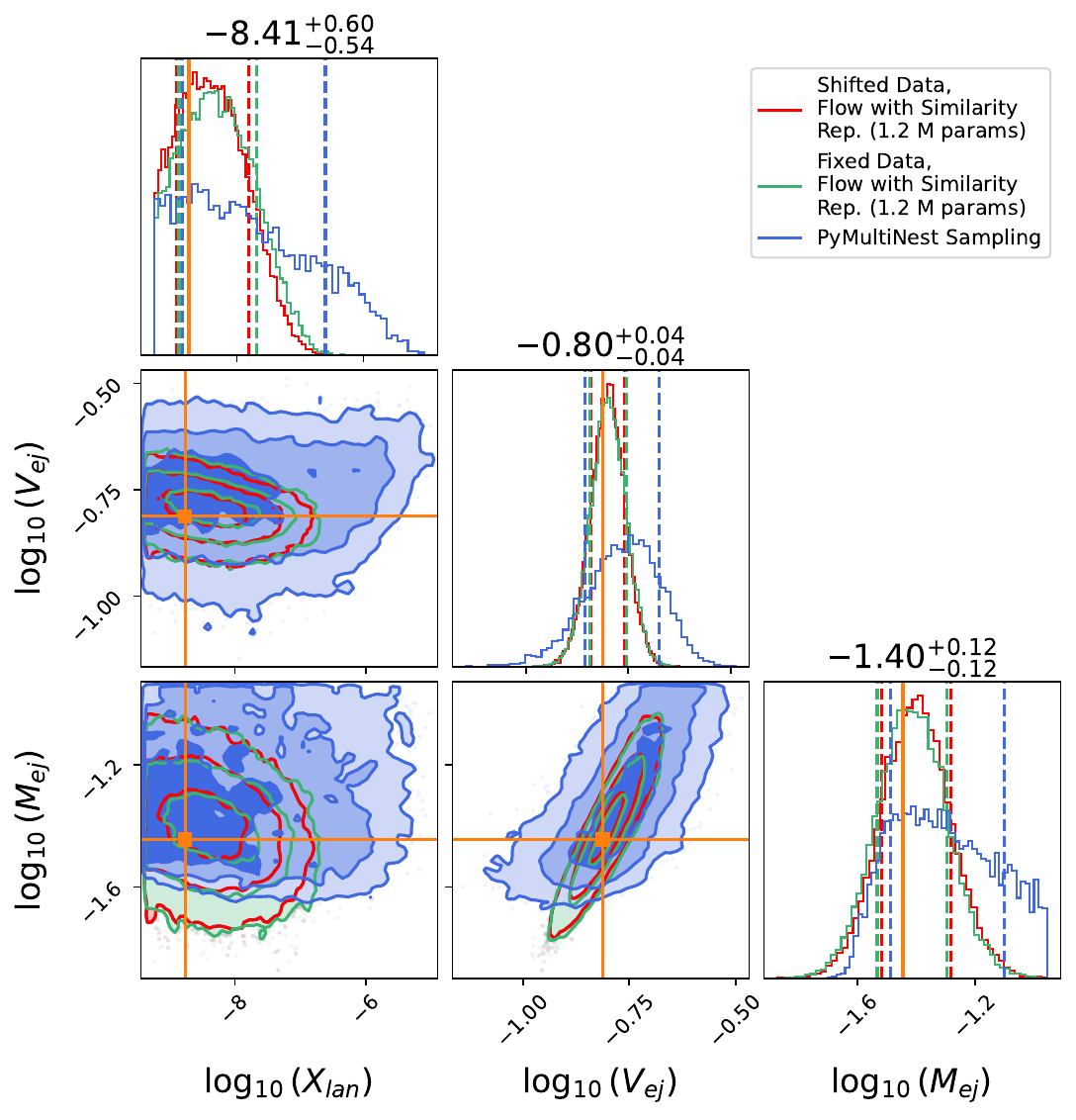}
\includegraphics[width = 8 cm]{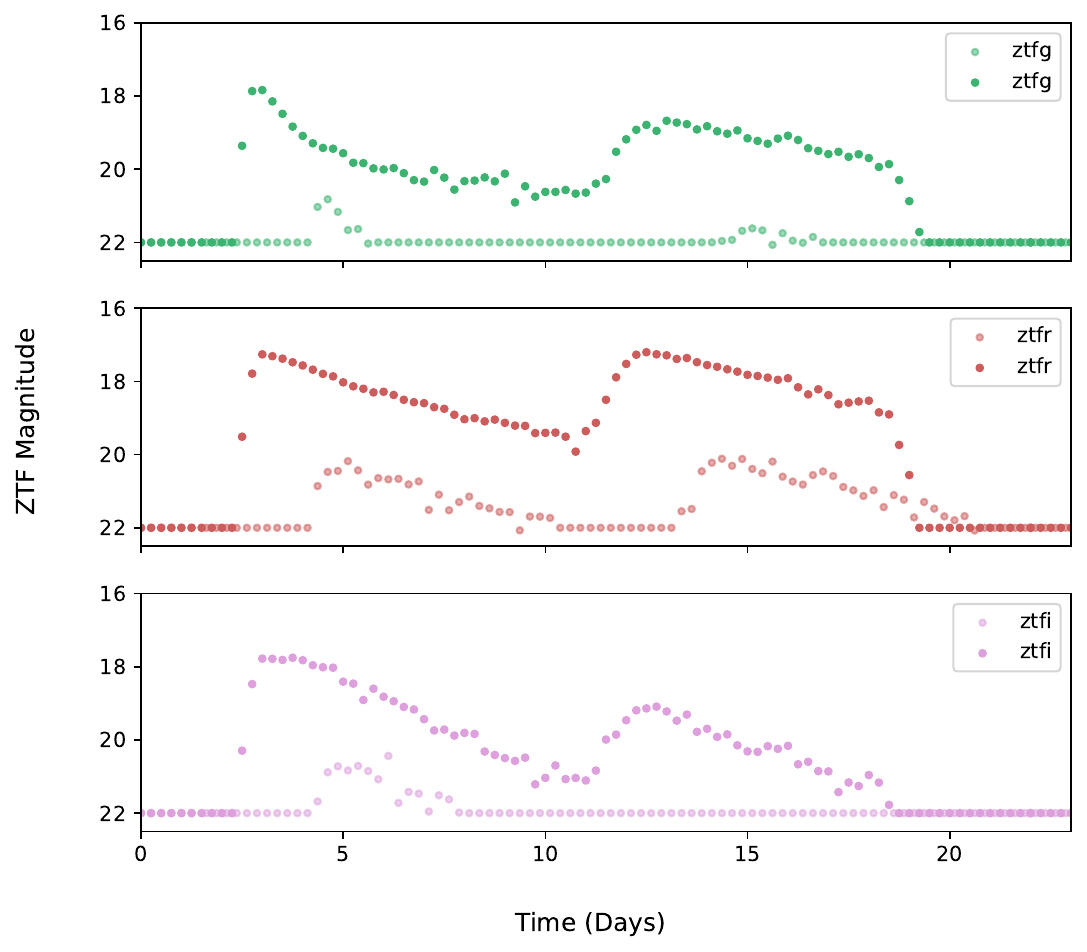}
\includegraphics[width = 8 cm]{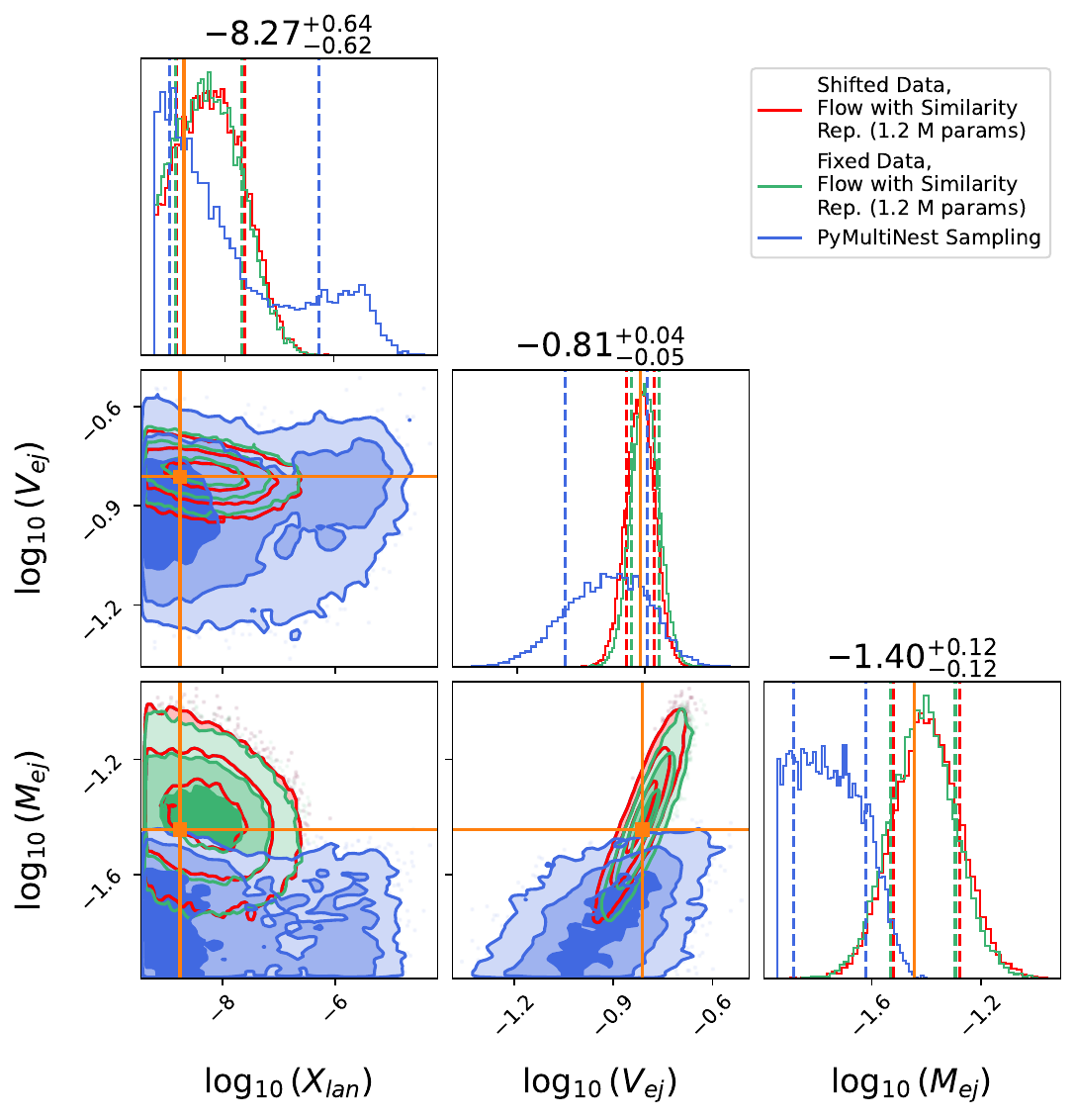}
\caption{Left panels: Fixed and shifted light curves for injection parameters of \logmej $= -1.44$, \logvej $= -0.81$, and \logxlan $= -8.76$. Upper left: Shifted light curve with $d_L = 50$ Mpc. Bottom left: Shifted light curve with $d_L = 200$ Mpc. Right panels: Corresponding posterior distributions with the shifted light curve posterior from the flow shown in red, its corresponding fixed counterpart inferred from the flow shown in green, and a comparison to {\multinest} shown in blue. We see agreement between the fixed and shifted posteriors from the flow, and that the three inferred parameters are in agreement with the truth. We find that the flow is consistent regardless of the distance, while \multinest\ faces more difficulty constraining parameters at $d_L = 200$ Mpc.}
\label{fig:comppost}
\end{figure*}

\begin{figure*}
    \centering
    \includegraphics[width = 17.5 cm]{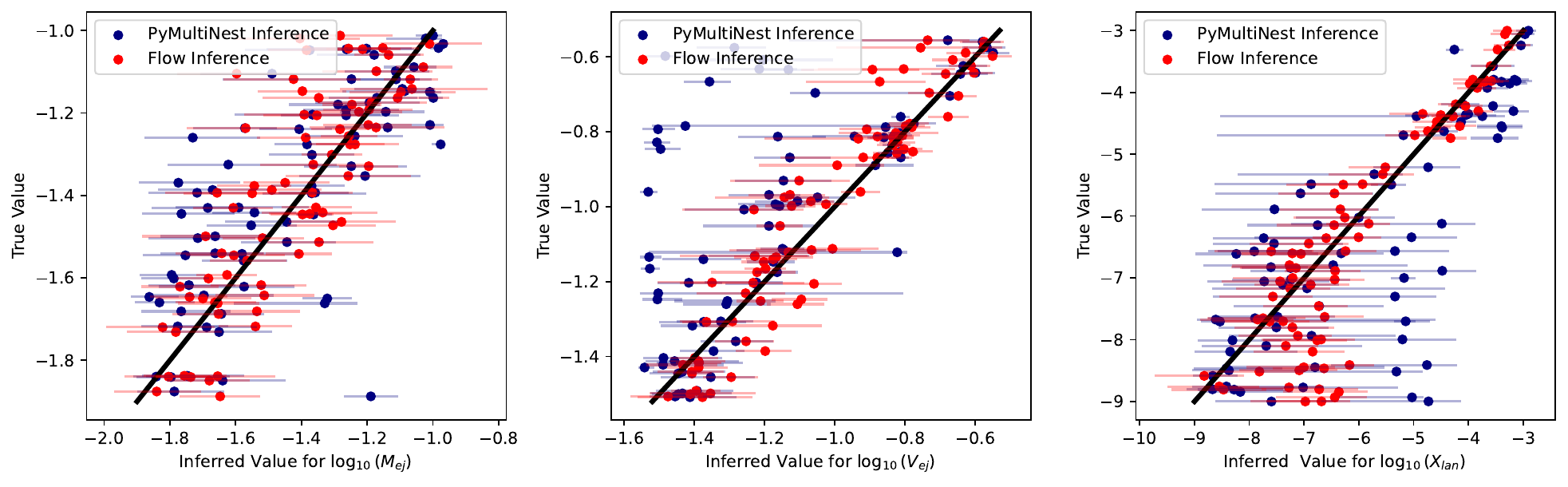}
    \caption{Plotted are 77 inferences and errors for {\mej}, {\vej}, and {\xlan} from \multinest\ in blue, and values from the normalizing flow in red. The inferred values of the parameter estimation methods are plotted against the true value of the parameter. The black line represents the behavior we expect where the inferred and true parameters are the same. Overall, our model performs well for the quantities $\log_{10}(M_{ej})$, $\log_{10}(V_{ej})$, and $\log_{10}(X_{lan})$. \multinest\ has occasional difficulties in constraining a value for $\log_{10}(V_{ej})$ and $\log_{10}(X_{lan})$ for the provided distributions, while our method has difficulty for high $\log_{10}(M_{ej})$. We utilize the identical shifted light curve sample used in the percentile-percentile plot in the lower panel of Figure~\ref{fig:flowloss} for these inferences.} 
    \label{fig:100inf}
\end{figure*}

We compare the performance of our model to nested sampling via the package {\multinest} \citep{2014A&A...564A.125B}. We maintain the same priors used for data generation, shown in Table~\ref{tab:dataprior}, while generating the posterior plots. To inform {\multinest} of the distribution, we use the same prior, but allow the parameter search space to remain $10\%$ larger than the original distribution to accommodate values at the edges of our priors. We assume the same priors for our model. The default sampler used by NMMA for light curve analysis is \multinest, thus we utilize this sampler for our comparison inferences.

We display our comparison of the fixed and shifted posteriors to {\multinest} sampling in Figure~\ref{fig:comppost} for one example injection of \logmej, \logvej, and \logxlan. A comparison is made between a signal at $d_L = 50$ Mpc and one at $d_L = 200$ Mpc, showing the difficulty nested sampling has when the light curve peak has shifted significantly. Additionally, distant light curves are dimmer and provide fewer data points for the sampling method to leverage. Our LFI method remains invariant to time, as displayed by the complete overlapping of the fixed and the shifted posteriors. The distributions of the flow posteriors also remain consistent for both luminosity distances, demonstrating distance invariance. 

To create a statistical sample of posteriors, we draw 100 combinations of our injection parameters from our uniform data distribution which serve as our test set. The light curve data is passed to {\multinest}, which uses 2048 live points to sample a posterior result. Due to an error in \multinest's sampling software that returns the same initial value for \logvej, 23 out of the 100 test cases are discarded from the sample. The error in \multinest's software is outside of the scope of this paper. We obtain posteriors from the flow utilizing the identical set of 77 shifted light curves. The resulting estimates are compared for our three parameters of interest in Figure~\ref{fig:100inf}.

We expect to see a linear correlation between the true values and the inferences. We observe that the flow performs well for all three parameters: \logmej, \logvej, and \logxlan. Certain areas of parameter space are notably challenging to accurately predict. In particular, small quantities of lanthanide can be misjudged by both the normalizing flow and by \multinest. High amounts of lanthanides have a distinct effect in the reddening of a kilonova light curve. As that amount reduces, the effect on the observed light curve is minimal. As a result, smaller values of $\log_{10}(X_{lan})$  are increasingly difficult to predict. We note that our recovery of $\log_{10}(V_{ej}/c)$ has an increased precision than \multinest, while the spread of $\log_{10}(M_{ej}/M_{\odot})$ is similar between the two methods.

\begin{table}
        \centering
	\caption{Accuracy of the two methods. We check whether the true value falls within the predicted result's range, defined by its error bars for both the flow and \multinest. We then calculate the percentage of the 77 results that contain the true value within the error.}
	\label{tab:accuracy}
	\begin{tabular}{lcc} 
		\hline\hline
		  Parameter & \multinest & LFI \\
		\hline
             \logmej & 62.34 \% & 66.23 \% \\
             \logvej & 58.44 \% & 64.94 \% \\
             \logxlan & 61.04 \% & 72.73 \% \\
             \hline\hline
	\end{tabular}
\end{table}

We compute the percentage of accurate inferences from both methods in Table \ref{tab:accuracy}. A result is deemed accurate if the truth falls within the associated error bars surrounding the estimate. Combining this information with the precision distributions in Figure \ref{fig:100inf}, we conclude that our method is competitive with \multinest\ in both accuracy and precision for all three parameters of interest.

\subsection{Model Resilience} \label{missingdata}

Though our model has the ability to infer parameters for a wide variety of kilonova light curves, our training dataset is limited by its regular cadence. To relax the requirement for rapid detections, we chose a cadence of six hours. 
Based on the recommendation of a rapid cadence between 40 seconds to 30 minutes for transient science from the LSST Science Requirements Document, our cadence selection is reasonable for kilonova detection \citep{lsststrategy}. However, we can leverage interpolation to test our model's resilience to missing or slowly sampled data. 

\begin{figure*}
    \centering
    \includegraphics[width = \columnwidth]{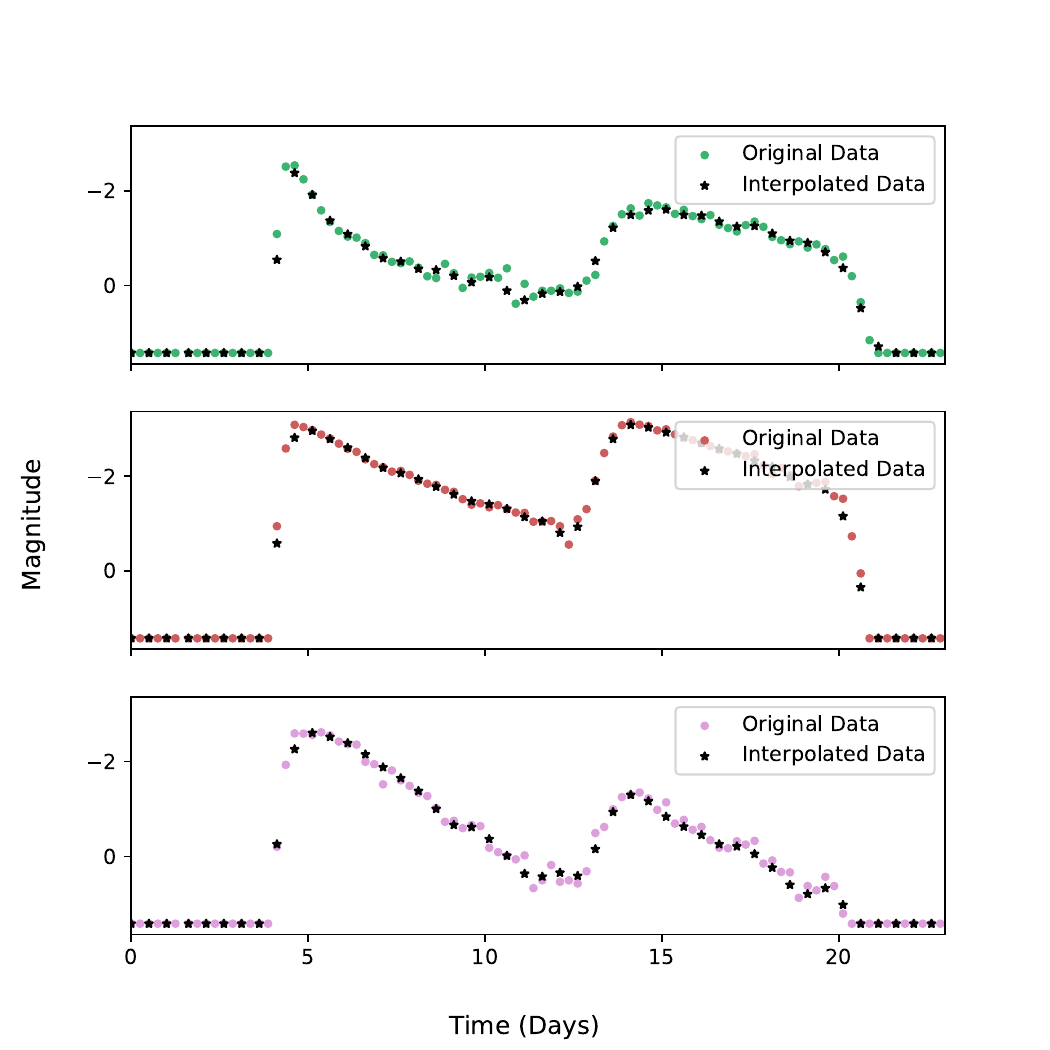}
    \includegraphics[width = \columnwidth]{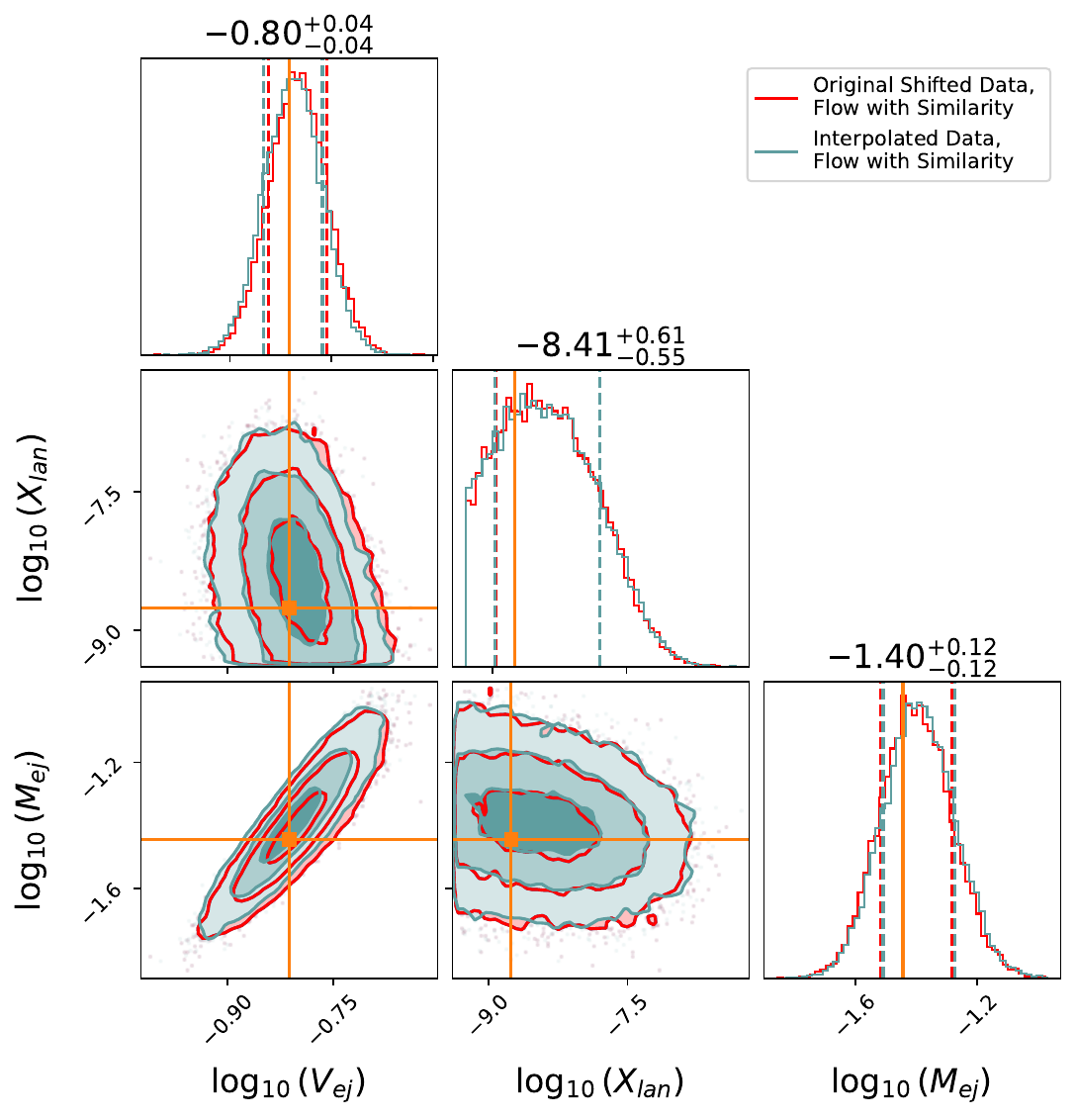}
    \caption{Left: Original light curve data plotted in color with the interpolated points overplotted as black stars for all three ZTF bands. Right: Posteriors for a light curve with \logmej $= -1.0417$, \logvej $= -1.3917$, and \logxlan $= -7.0289$. Blue represents the flow results on the original data, and yellow with every other data point interpolated. The two posteriors overlap, indicating that interpolating these missing values has little effect on the resulting distribution.}
    \label{fig:everyotherinterp}
\end{figure*}

As we intend for our framework to be used with a variety of optical instruments, it is imperative that the model can adjust to several cadences. For our first test,  we discard every other detection and linearly interpolate new data points in their stead. We find that the model has sufficient information to generate an inference with high accuracy. Figure~\ref{fig:everyotherinterp} displays a sample posterior of the interpolation in grey with the original in red alongside the respective interpolated light curve. We observe alignment between the two posteriors. This test indicates that our model has some flexibility to accommodate different observational cadences to perform accurate parameter estimation.

We expect observational light-curve data to be observed at a nonuniform cadence, and it is possible that key points near the peak will go undetected. Data can also be interpolated through the use of transformers to achieve uniform cadences from nonuniform detections in the case where linear interpolation is not adequate \citep{sharma2024}. Similar work on analyzing supernova light curves has utilized Gaussian Process fitting to interpolate unevenly sampled light curves as a data pre-processing step prior to using normalizing flows \citep{2022arXiv221104480V}. These methods provide avenues for further investigation into the creation of a more robust model for telescopic data.

For further tests in lower cadence regimes, see Appendix~\ref{app:resilience} for details on using interpolation to replace missing portions of the data. 

\section{Conclusion} \label{Conclusion}

We present a method to perform parameter estimation on kilonova lightcurves using LFI. Our framework provides a parameter estimation method competitive with pre-existing samplers. We demonstrate that self-supervised learning successfully marginalizes our invariant parameters of distance and time, decreases the dimensionality of our light curve data, and allows the normalizing flow to converge faster. Additionally, our model provides posteriors in less than a second, while {\multinest} requires time on the order of minutes. This allows our model to be a viable alternative inference method that has the added benefit of time and distance invariance.  

Missing data is a common problem that can limit many ML approaches to classification and inference. We demonstrate that for a full day of missing data, our model returns statistically similar inferences when linear interpolation is applied. Further investigation into utilizing observationally realistic light curves will be required to launch this model for analysis on data from Rubin.

Our method is constrained by the choice of kilonova model. In general, light curve models are parametrized by different physical inputs. For example, the model described in \cite{2023MNRAS.520.2558B} allows for priors on wind, inclination angle, and the electron fraction of the neutron star in addition to the ejecta mass and velocity parameters found in the Ka2017 model. Including these physical parameters requires retraining and slight restructuring to our network. However, the approach remains consistent regardless of model choice. 

Similarly, our model is fine-tuned for ZTF bands g, r, and i. The model requires information for at least one of these bands, but can be retrained for any number of bands from any instruments provided that the corresponding data can be simulated. As we transition between instruments for light curve observation, our model remains flexible to training with data from multiple bands. The Vera C. Rubin Observatory currently plans to optimize simultaneous light curve sampling in multiple bands, including g, r, and i bands that largely overlap with ZTF \citep{2019ApJ...873..111I, 2019PASP..131a8002B}. Multiple filters can be added or removed by changing the number of data channels supplied to our data summary, $\Gamma$. Through a retraining of the model, our framework can be utilized for any instrument. 

The use of self-supervised learning as a marginalization technique and normalizing flows to perform LFI can also be applied to other light curve parameters. Previous work has investigated the role of machine learning techniques on supernova light curves. For example, the \texttt{Superphot+} classifier uses a gradient-boosted machine architecture to assign types to ZTF supernova light curves \citep{2024arXiv240307975D}. \texttt{Superphot+} relies on nested sampling to estimate several light curve parameters. With a novel dataset, our architecture could be applied to improve secondary classifiers by providing fast and accurate parameter estimation. Since we demonstrate flexibility in cadences and missing data in Section~\ref{missingdata} and Appendix~\ref{app:resilience}, our framework is capable of being applied to observational data. 

Our framework is publicly available and ready to use through the NMMA package. With the given constraints mentioned here, a user can access this inference method through the command line and produce a posterior plot for a given NMMA-generated light curve or injection file. Future work is being done to improve the parameter estimation of our method, and to retrain with a separate light curve simulation models and filters to create a host of network weights that can return posteriors for a broad range of kilonova light curves. 

\section*{Acknowledgements}

We thank the NMMA team for their invaluable discussions and suggestions for additional testing of the model framework and the incorporation of the model into the NMMA codebase. We thank  Peter Pang for reviewing the code. We thank Ethan Marx, Will Benoit, and Alec Gunny for their helpful comments. This work is supported by NSF HDR Institute Grant PHY-2117997, “Accelerated AI Algorithms for Data-Driven Discovery.” M.W.C also acknowledges support from the National Science Foundation with grant numbers PHY-2308862 and PHY-2409481. We thank Michelle Ntampaka of the Space Telescope Science Institute for insightful discussions into building robust neural networks.

%%%%%%%%%%%%%%%%%%%%%%%%%%%%%%%%%%%%%%%%%%%%%%%%%%
\section*{Data Availability}

The data processing code, neural network and normalizing flow code, training code, and model weights have been integrated into the NMMA codebase at: \url{https://github.com/nuclear-multimessenger-astronomy/nmma}. Through this package, command line inference using LFI is available. Previous and ongoing code development is located at: \url{https://github.com/malinadesai/Kilo}. The training data is also available for use at: \url{https://doi.org/10.5281/zenodo.11640528}.

%%%%%%%%%%%%%%%%%%%% REFERENCES %%%%%%%%%%%%%%%%%%

\bibliographystyle{mnras}
\bibliography{main} 

%%%%%%%%%%%%%%%%%%%%%%%%%%%%%%%%%%%%%%%%%%%%%%%%%%

%%%%%%%%%%%%%%%%% APPENDICES %%%%%%%%%%%%%%%%%%%%%

\appendix

\section{Hyperparameter Tuning} \label{app:tuning}

\begin{table*}
	\centering
	\caption{Ten best hyperparameter tuning configurations for the similarity embedding parameters.}
	\label{tab:hypeembed}
	\begin{tabular}{lccccccccr}
		\hline\hline
		Trial & Dimensions & Kernel Size & $h$ Layers & Blocks & Final Dimension & Batch Size & Learning Rate & Final Train Loss & Final Val. Loss \\
		\hline
            2 & 7 & 5 & 1 & 4 & 5 & 50 & 2.75e-05 & 0.01502 & 0.01172 \\
            3 & 7 & 7 & 1 & 4 & 5 & 25 & 1.32e-05 & 0.03583 & 0.12160 \\
            7 & 5 & 7 & 3 & 6 & 2 & 25 & 3.51e-05 & 0.02391 & 0.02914 \\
            10 & 8 & 7 & 1 & 6 & 2 & 25 & 2.64e-05 & 0.01324 & 0.03076 \\
            11 & 5 & 5 & 1 & 3 & 2 & 25 & 4.31e-06 & 0.02485 & 0.13644 \\
            17 & 5 & 9 & 1 & 5 & 2 & 25 & 1.67e-05 & 0.01662 & 0.13725 \\
            31 & 5 & 7 & 2 & 6 & 2 & 25 & 4.71e-06 & 0.01231 & 0.02960 \\
            40 & 8 & 9 & 2 & 3 & 2 & 50 & 2.39e-05 & 0.04336 & 0.05734 \\
            48 & 8 & 9 & 1 & 5 & 2 & 50 & 8.60e-05 & 0.01128 & 0.02089 \\
            98 & 4 & 7 & 1 & 4 & 5 & 50 & 1.13e-06 & 0.05847 & 0.10924 \\
		\hline\hline
	\end{tabular}
\end{table*}

\begin{table*}
	\centering
	\caption{Ten best hyperparameter tuning configurations for the normalizing flow parameters.}
	\label{tab:hypeflow}
	\begin{tabular}{lccccccr}
		\hline\hline
		Trial & Blocks & Transforms & Hidden Features & Batch Size & Learning Rate & Final Train Loss & Final Val. Loss \\
		\hline
            0 & 8 & 5 & 100 & 100 & 6.39e-05 & 0.58579 & 0.55771 \\
            4 & 10 & 4 & 100 & 25 & 3.38e-05 & 0.56152 & 0.60114 \\
            11 & 9 & 7 & 80 & 25 & 4.09e-05 & 0.65374 & 0.72734 \\
            19 & 6 & 5 & 70 & 25 & 4.05e-05 & 0.71651 & 0.82130 \\
            28 & 10 & 6 & 100 & 25 & 8.76e-05 & 0.34584 & 0.33634 \\
            40 & 9 & 5 & 90 & 25 & 9.12e-05 & 0.09280 & -0.18805 \\
            57 & 6 & 5 & 90 & 25 & 9.28e-05 & 0.37192 & 0.53521 \\
            65 & 9 & 6 & 70 & 25 & 3.82e-05 & 0.79535 & 0.90459 \\
            75 & 10 & 7 & 70 & 25 & 6.89e-5 & 0.37713 & 0.40245 \\
            98 & 9 & 6 & 80 & 25 & 9.33e-05 & 0.12195 & 0.04436 \\
		\hline\hline
	\end{tabular}
\end{table*}

\begin{figure*} 
\centering
    \includegraphics[width = \columnwidth]{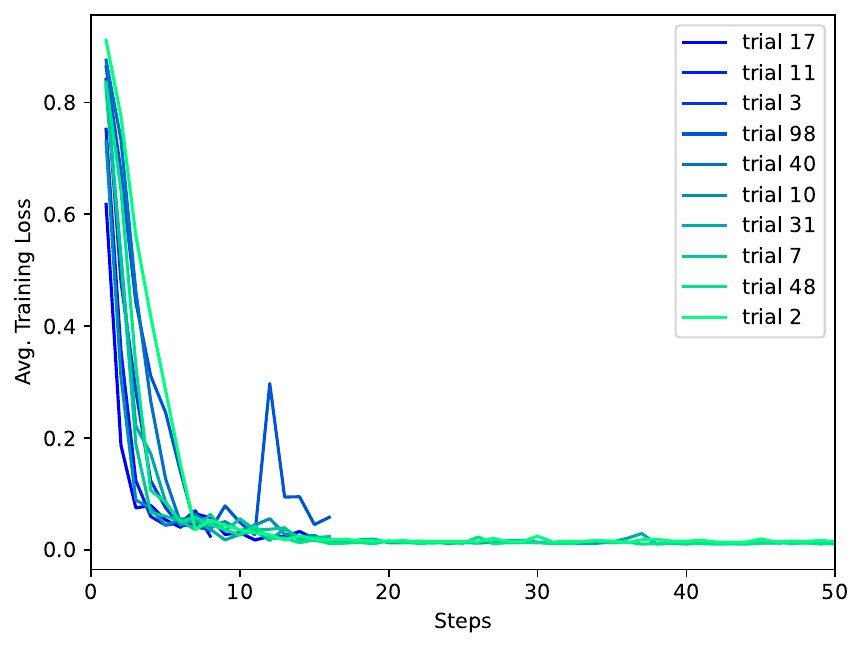}
    \includegraphics[width = \columnwidth]{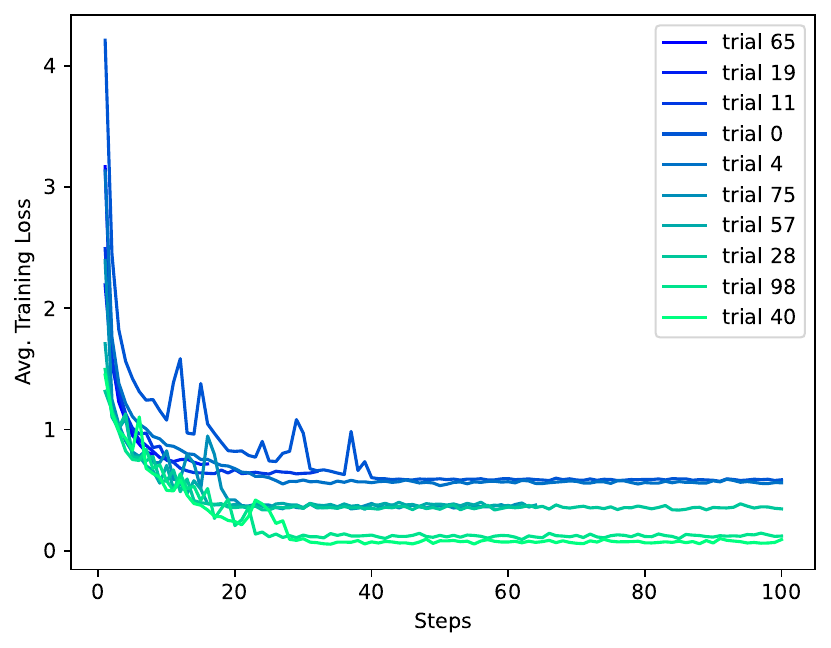}
    \caption{Training loss curves for the 10 best configurations of the embedding network (upper) and the normalizing flow (lower). We utilize the configurations of the network parameters selected in trial 2 and trial 40 for our models as they converge to the lowest training and validation loss.}
    \label{fig:tuning}
\end{figure*}

Tables \ref{tab:hypeembed} and \ref{tab:hypeflow} display the ten best hyperparameter tuned runs for the embedding network and the normalizing flow. Trial 2 and Trial 40 are selected as the respective best trials based on the lowest validation loss achieved at the termination of the run. These runs are shown in Figure~\ref{fig:tuning}, which displays the training loss curves for the ten best trials for the embedding and for the normalizing flow. To increase the efficiency of hyperparameter searches, we leverage the Async HyperBand Scheduler (ASHA) from \texttt{Ray}. This scheduling algorithm uses the user-selected optimization metric (here, minimizing the validation loss) to identify poorly performing runs and terminate them early \citep{li2018massively}. In this way, we can efficiently search over the parameter spaces in Table~\ref{tab:config} for our models. 

\section{Further Resilience Testing} \label{app:resilience} 

\begin{figure*}
    \centering
    \includegraphics[width = \columnwidth]{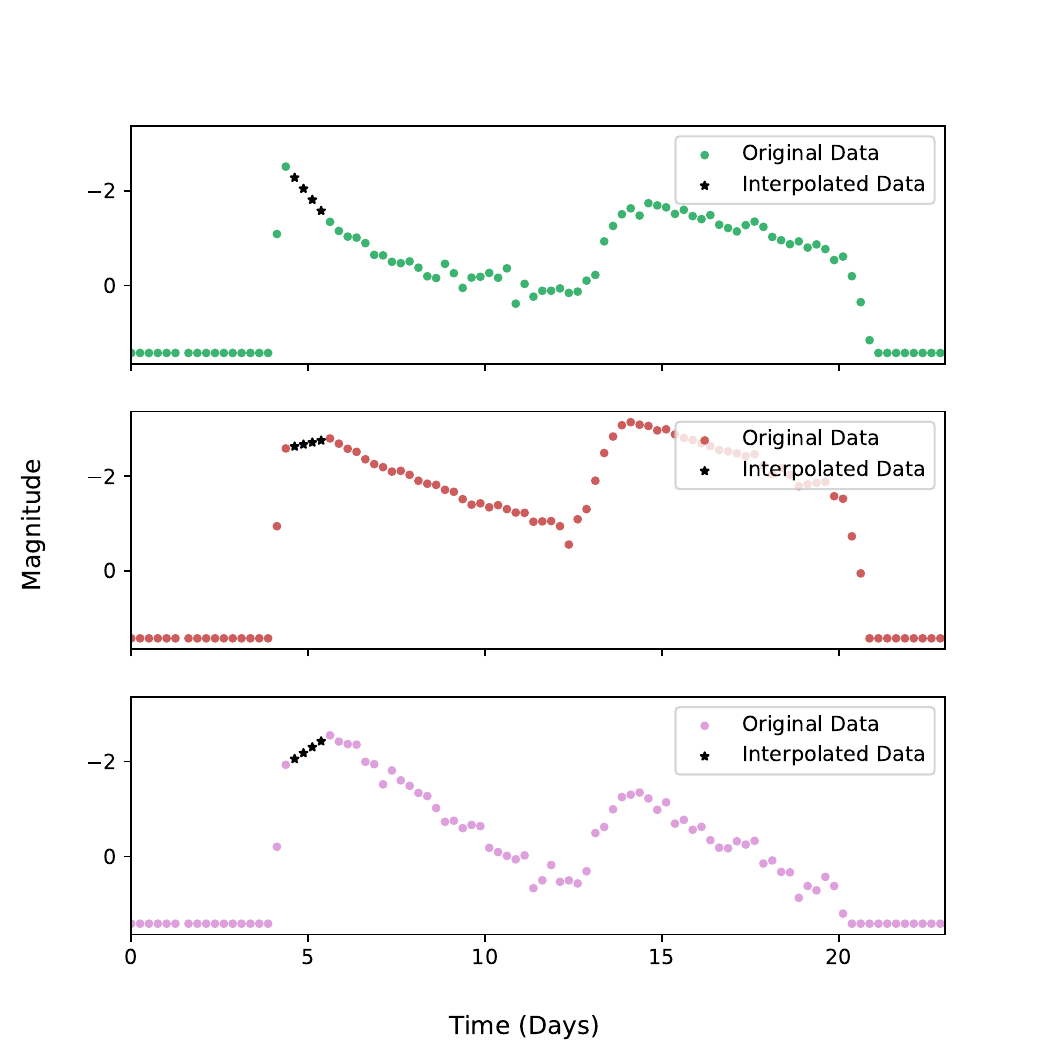}
    \includegraphics[width = \columnwidth]{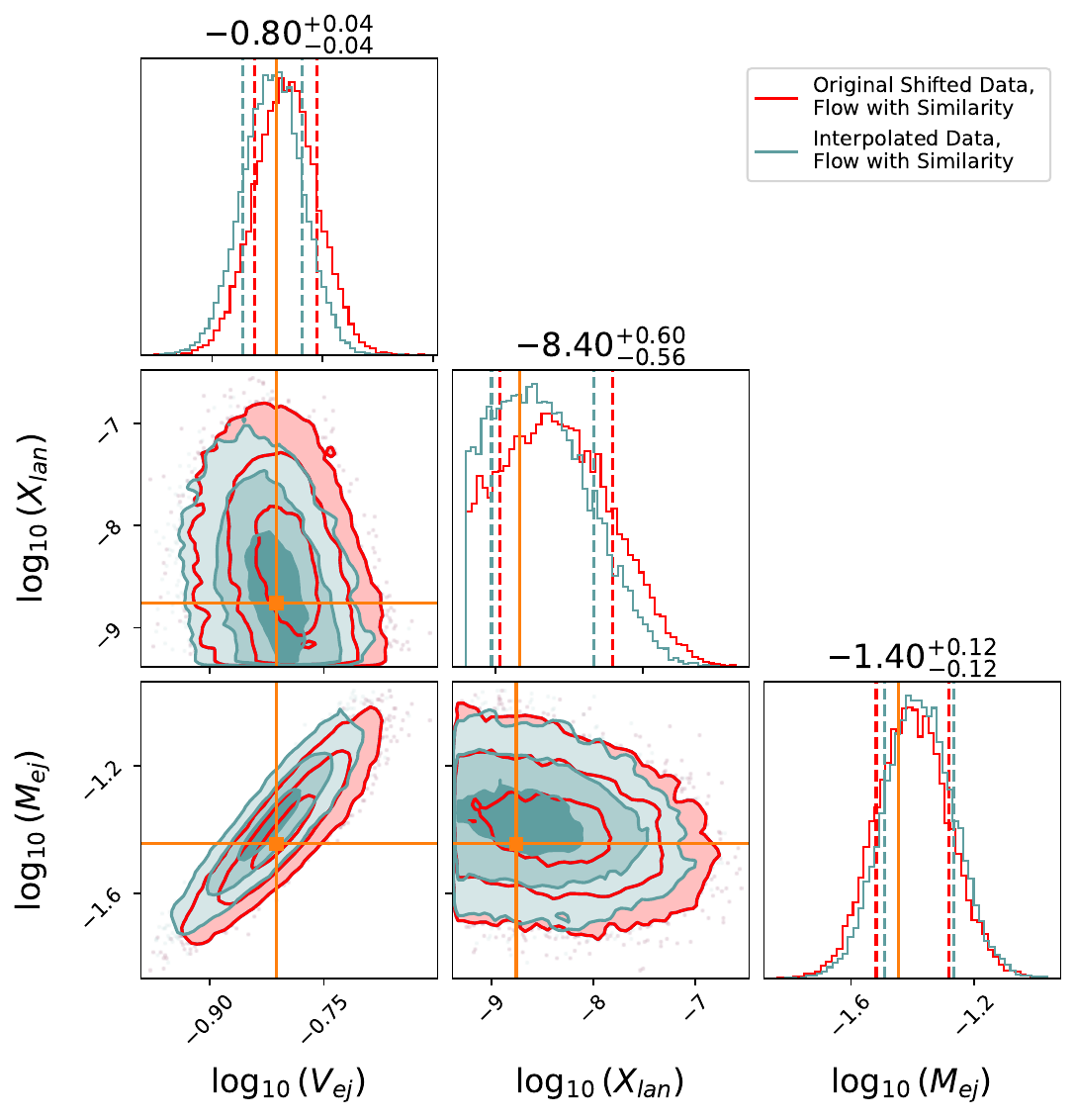}
    \caption{Left: Original light curve data plotted in color with the interpolated points plotted on top as black stars for all three ZTF bands. Right: Posteriors for the light curve pictured, with injection parameters \logmej $= -1.44$, \logvej $= -0.81$, and \logxlan $= -8.76$. Red represents the flow results on the original shifted light curve data, and grey represents the flow results for light curve data with four consecutive data points interpolated. The two posteriors mostly overlap, indicating that interpolating  missing values has a minor effect on the resulting distribution.}
    \label{fig:interppost}
\end{figure*}

We investigated the results of interpolation to generate missing sections of data by removing four consecutive data points from each band near the peak and performing a linear interpolation over a 36-hour time span. The posteriors for the original and the interpolated light curves are well aligned. A sample case is shown in Figure~\ref{fig:interppost} with the interpolated light curve shown on the left and its posterior shown in grey on the right. The original light curve posterior is shown in red. The two posteriors are aligned, providing a good indication that linear interpolation can be sufficient when the data has missing values. We also highlight that the true values of the light curve are correctly predicted by the model. Further work into interpolation schemes will be a useful avenue to increasing the accuracy of predicted points.

%%%%%%%%%%%%%%%%%%%%%%%%%%%%%%%%%%%%%%%%%%%%%%%%%%
\bsp	
\label{lastpage}
\end{document}